    \documentclass{bmcart}
    
    \usepackage{amsthm,amsmath}
    \RequirePackage{hyperref}
    \usepackage[utf8]{inputenc} %unicode support
    \usepackage{booktabs}
    \usepackage{enumerate}
    \usepackage{graphics}
    \usepackage{adjustbox}
   
    \usepackage{hyperref}
    \usepackage{cleveref}
    \crefformat{footnote}{#2\footnotemark[#1]#3}
    \usepackage[titletoc,title]{appendix}
    
    \newcommand{\add}[1]{\textcolor{black}{#1}}
    \startlocaldefs
    \endlocaldefs
    
\begin{document}
    
    %%% Start of article front matter
    \begin{frontmatter}
    
    \begin{fmbox}
    \dochead{Research}
    
    
    \add{\title{Following the footsteps of giants: Modeling the mobility of historically notable individuals using Wikipedia}}
    
    %%%%%%%%%%%%%%%%%%%%%%%%%%%%%%%%%%%%%%%%%%%%%%
    %%                                          %%
    %% Enter the authors here                   %%
    %%                                          %%
    %% Specify information, if available,       %%
    %% in the form:                             %%
    %%   <key>={<id1>,<id2>}                    %%
    %%   <key>=                                 %%
    %% Comment or delete the keys which are     %%
    %% not used. Repeat \author command as much %%
    %% as required.                             %%
    %%                                          %%
    %%%%%%%%%%%%%%%%%%%%%%%%%%%%%%%%%%%%%%%%%%%%%%
    
    \author[
       addressref={aff1, aff2},                   % id's of addresses, e.g. {aff1,aff2}
       corref={aff1},                       % id of corresponding address, if any
       email={llucchini@fbk.eu}   % email address
    ]{\inits{LL}\fnm{Lorenzo} \snm{Lucchini}}
\author[
       addressref={aff1},
       email={satonelli@fbk.eu}
    ]{\inits{ST}\fnm{Sara} \snm{Tonelli}}
\author[
       addressref={aff1},
       email={lepri@fbk.eu}
    ]{\inits{BL}\fnm{Bruno} \snm{Lepri}}
    
    %%%%%%%%%%%%%%%%%%%%%%%%%%%%%%%%%%%%%%%%%%%%%%
    %%                                          %%
    %% Enter the authors' addresses here        %%
    %%                                          %%
    %% Repeat \address commands as much as      %%
    %% required.                                %%
    %%                                          %%
    %%%%%%%%%%%%%%%%%%%%%%%%%%%%%%%%%%%%%%%%%%%%%%
    
    \address[id=aff1]{%                           % unique id
      \orgname{Fondazione Bruno Kessler}, % university, etc
      \street{via Sommarive, 18},                     %
      \postcode{38122}                                % post or zip code
      \city{Trento},                              % city
      \cny{Italy}                                    % country
    }
    \address[id=aff2]{%                           % unique id
      \orgname{Department of Information Engineering and Computer Science, University of Trento}, % university, etc
      \street{via Sommarive, 9},                     %
      \postcode{38122}                                % post or zip code
      \city{Trento},                              % city
      \cny{Italy}                                    % country
    }

    %%%%%%%%%%%%%%%%%%%%%%%%%%%%%%%%%%%%%%%%%%%%%%
    %%                                          %%
    %% Enter short notes here                   %%
    %%                                          %%
    %% Short notes will be after addresses      %%
    %% on first page.                           %%
    %%                                          %%
    %%%%%%%%%%%%%%%%%%%%%%%%%%%%%%%%%%%%%%%%%%%%%%
    
    % \begin{artnotes}
    % %\note{Sample of title note}     % note to the article
    % \note[id=n1]{Equal contributor} % note, connected to author
    % \end{artnotes}
    
    \end{fmbox}% comment this for two column layout
    
    %%%%%%%%%%%%%%%%%%%%%%%%%%%%%%%%%%%%%%%%%%%%%%
    %%                                          %%
    %% The Abstract begins here                 %%
    %%                                          %%
    %% Please refer to the Instructions for     %%
    %% authors on http://www.biomedcentral.com  %%
    %% and include the section headings         %%
    %% accordingly for your article type.       %%
    %%                                          %%
    %%%%%%%%%%%%%%%%%%%%%%%%%%%%%%%%%%%%%%%%%%%%%%
    
    \begin{abstractbox}
    \begin{abstract} % abstract
    %\parttitle{First part title} %if any
    The steady growth of digitized historical information is continuously stimulating new different approaches to the fields of \add{Digital Humanities and Computational Social Science}. In this work we use Natural Language Processing techniques to retrieve large amounts of historical information from Wikipedia. In particular, the pages of a set of \add{historically notable individuals} are processed to catch the locations and the date of people's movements. This information is then structured in a geographical network of mobility \add{patterns}. 

    We analyze the mobility of \add{historically notable individuals} from different perspectives to better understand the role of migrations and international collaborations in the context of innovation and cultural development. In this work, we first present some general characteristics of the dataset from a social and geographical perspective. Then, we build a spatial network of cities, and we model and quantify the tendency to explore of a set of people that can be considered as \add{historically and culturally notable}.
    In this framework, we show that by using a multilevel radiation model for human mobility, we are able to catch important features of migration's behavior. Results show that the choice of the target migration place for \add{historically and} culturally relevant people is limited to a small number of locations and that it depends on the discipline a notable is interested in \add{and on the number of opportunities she/he can find there}.
    %\parttitle{Second part title} %if any
    %Text for this section.
    \end{abstract}
    
    %%%%%%%%%%%%%%%%%%%%%%%%%%%%%%%%%%%%%%%%%%%%%%
    %%                                          %%
    %% The keywords begin here                  %%
    %%                                          %%
    %% Put each keyword in separate \kwd{}.     %%
    %%                                          %%
    %%%%%%%%%%%%%%%%%%%%%%%%%%%%%%%%%%%%%%%%%%%%%%
    
    
    % MSC classifications codes, if any
    %\begin{keyword}[class=AMS]
    %\kwd[Primary ]{}
    %\kwd{}
    %\kwd[; secondary ]{}
    %\end{keyword}
    
    \end{abstractbox}
    %
    %\end{fmbox}% uncomment this for twcolumn layout
    
    \end{frontmatter}
    
    %%%%%%%%%%%%%%%%%%%%%%%%%%%%%%%%%%%%%%%%%%%%%%
    %%                                          %%
    %% The Main Body begins here                %%
    %%                                          %%
    %% Please refer to the instructions for     %%
    %% authors on:                              %%
    %% http://www.biomedcentral.com/info/authors%%
    %% and include the section headings         %%
    %% accordingly for your article type.       %%
    %%                                          %%
    %% See the Results and Discussion section   %%
    %% for details on how to create sub-sections%%
    %%                                          %%
    %% use \cite{...} to cite references        %%
    %%  \cite{koon} and                         %%
    %%  \cite{oreg,khar,zvai,xjon,schn,pond}    %%
    %%  \nocite{smith,marg,hunn,advi,koha,mouse}%%
    %%                                          %%
    %%%%%%%%%%%%%%%%%%%%%%%%%%%%%%%%%%%%%%%%%%%%%%
    
    %%%%%%%%%%%%%%%%%%%%%%%%% start of article main body
    % <put your article body there>
    
    %%%%%%%%%%%%%%%%
    %% Background %%
    %%

\section{Introduction}

Ever since the first villages were built by primitive people, humankind has moved from one community to another, in search of better life conditions or new opportunities \cite{manning2005}. These conditions can be represented by different factors, such as the opportunity of a job, better living standards, or the distance from the home country \cite{lee1966,arango2000}. The set of all these factors is difficult to define a-priori. 

For example, according to the dominant neo-classical theory, people tend to make choices in order to maximize their income or level of well-being \cite{migration,todaroSmith2006}. Thus, the search for better economic conditions is one of the most important factors in the decision of moving from one location to a more attractive one. However, there are other factors that could play an important role in the decision-making process of a specific group of people. The attractiveness of an opportunity can also depend on  cultural and linguistic barriers, and on the presence of particular communities at the destination \cite{migration,guiso2009,belotEderveen2012}. Hence, in order to define a complete migration framework it is important to consider all the linked aspects that intervene in modifying the attractiveness of a location (e.g. economical, environmental, cultural, and political aspects) \cite{migration,belotEderveen2012,beine2015}. As a consequence, we believe that, in the same way as the economic conditions play a central role for those seeking employment, it is very relevant to investigate if there are other factors playing a role for specific kinds of migration, for example the migration of notable people and intellectuals in the course of history \cite{khoshkish1966}.

For the general problem of human migration different mathematical models have been built both in a descriptive \cite{gravity,intervening,utilities,radiation} and predictive \cite{barlacchi} framework. However, these models have never been applied to the specific scenario of modeling the historical movements and migrations of intellectual figures. Instead, mobility and migrations historically played, and still play, an important role in the process of cultural evolution introducing seeds of change in different places around the world \cite{cavalli-sforza81book}. Thus, understanding the patterns that \add{historically notable individuals} followed during their lives and how these affected the cultural evolution \add{and the human history} is an intriguing and still open research question \cite{grandchallenges}. This approach introduces new challenges because of the specificity of the problem and the relatively small number of people that had an impact on cultural evolution \add{and human history}. New perspectives were opened by a recent approach proposed by Schich \emph{et al.} \cite{culture}. By analyzing birth and death locations \add{historically notable individuals} they captured, from a network perspective, the key characteristics of their exploratory behavior. 

In our paper, we propose a way of using Natural Language Processing (NLP) and Network Science techniques to further characterize and model the mobility of \add{historically and culturally notable individuals} and to investigate the factors playing a role in their migration patterns.

In the last decade, the NLP community has developed technologies for extracting information from unstructured texts, thus enabling their application also to  interdisciplinary research areas. 
Understanding and modeling historical migration phenomena require specific historical data. To this end, we propose to use NLP techniques to process the digital biographies contained in Wikipedia and to extract migratory events from its encyclopedic information. In particular, a subset of the biographies available in the English version of Wikipedia is used as raw data source. From this source, for each notable person \add{we search for her/his footsteps hidden in her/his Wikipedia page and collect} the following information: (i) the place and date of birth, (ii) the place and date of death, and (iii) the place and date of the various in-life migrations (e.g. moving from one city to a different one).
This results in a more complete set of data with respect to \cite{culture}, enriching the global picture of the mobility \add{of notable individuals} with a finer temporal granularity, while enabling, to the best of our knowledge, to model this process from a historical point of view for the first time.

As proposed in \cite{pantheon_yu}, we use the term \textit{culture} to focus our attention on the set of notable contributions to the development of human history in its broadest sense: from poetry to sports, from music to physics and mathematics. In particular, we model the mobility dynamics of notable people, namely those people whose cultural production is known at a global level. To this end, we introduce a modification to the \textit{radiation model} for human mobility \cite{radiation}. The assumption behind the \textit{radiation model} resides in using the city size (i.e. the city population) as a proxy of the number of job opportunities. In our paper, we modify the \textit{radiation model} to take into account, in addition to the role played by the city size, also the attractive role played by the different disciplines and by the number of notable people as proxies of cultural opportunities.

Then, we compare the predictive performances of our \textit{cultural-based radiation model} and of the state-of-the art population-based radiation model on three main aspects of the migration processes of notable people: (i) the radius of gyration of each notable person, (ii) the number of different cities the notable people lived in during their lives, and (iii) the distances between the source of a migration and its destination (jump lengths). Interestingly, our results show that the radius of gyration and the jump lengths are best modeled considering three different factors: (i) the population of the city, as a proxy for the economical wealth and job opportunities, (ii) the number of \add{notable individuals} that spent some time of their lives in a given city, as a proxy for the role played by the city as a cultural attractor, and (iii) the specific discipline an \add{historically and culturally notable person} is working in, as a proxy both for the interests a city has in investing on a specific cultural area and for the tendency that people, interested or working on that discipline, have to follow notable figures from the same domain.

Our results pave the way for further investigations on the historical role played by places and cities (e.g. ancient Athens, Renaissance Florence, Song Dynasty Hangzhou, Vienna of 1900, Silicon Valley, etc.) in becoming cultural attractors for \add{historically and culturally notable individuals} and, thus, making them grow into flourishing places for novel artistic and literary movements, for new scientific and philosophical theories, for social, technological and political innovations \cite{weiner2016}. 

\section{Materials and Methods} \label{materials_and_methods}
\subsection{Data}
\add{In this section, we present and} discuss how we build a dataset containing biographical information on thousands of historically relevant people in the context of cultural production and innovation. Together with information about their field of influence, we also extract information about their mobility patterns.

We start from the set of notable people identified by the Pantheon project \cite{pantheon_yu}. This project collected the biographies of $11,341$ \add{historically and} culturally relevant people that lived from $3500$ B.C. to $2010$ A.D. More precisely, the dataset was built by extracting information from the Wikipedia biography and info-boxes. Here, a person is defined as \textit{notable} if the corresponding Wikipedia page is translated in 25 or more different languages, since the focus is on global \add{historical and} cultural contributions.

For each notable person, several features were annotated and manually verified, among which we consider the following ones:
\begin{itemize}
    \item birth place (geo-localized), state, and date;
    \item occupation, work area, and discipline.
\end{itemize}

\begin{table}
    \centering
	\caption{Composition of the selected subset of notable people, organized by the different work areas and disciplines present in the Pantheon dataset \cite{pantheon_yu}. In particular, 15 work areas are presented, sorted by  discipline.}
\vspace{0.5cm}
	\begin{tabular}{ccr}
	\toprule
		Work Area & Discipline & Percentage (\%)\\
	\midrule
		Film and Theatre & Arts & 10.35\\
		Music & Arts & 8.81\\
		Fine Arts & Arts & 5.61\\
		Design & Arts & 1.68\\
	\midrule
		Natural Sciences & Science and Technology & 13.74\\
		Social Sciences & Science and Technology & 3.92\\
		Medicine & Science and Technology & 2.25\\
		Math & Science and Technology & 2.04\\
	\midrule
		Language & Humanities & 17.39\\
		Philosophy & Humanities & 3.51\\
	\midrule
		Government & Institutions & 15.90\\
		Military & Institutions & 3.81\\
	\midrule
		Activism & Public Figure & 1.23\\
	\midrule
		Individual Sports & Sports & 1.28\\
	\midrule
		Business & Business and Law & 1.07\\
		\bottomrule
	\end{tabular}\label{tab1}
\end{table}

While the process of selection of notable people in Yu \emph{et al.} \cite{pantheon_yu} resulted in $11,341$ biographies selected from more than one million biographies available in Wikipedia, we further restrict our analyses on the notable people that were actively working and producing cultural outcomes in a specific time-window, the first fifty years of the 20th century. The focus on this specific time-window is justified by two combined needs. The first is to have a sufficiently short time span so that the properties of the global mobility do not change.
The second is to have the highest possible number of notable people with complete and precise migratory information, i.e. with birth, in-life, and death locations. The first requirement reduces the width of the time-window, while the second one requires to consider relatively recent years. 
We therefore consider active the notable people that were at least 20 years old by the end of the considered time-window to ensure that their \add{historical and} cultural contribution  was made during this time span, and that no bias in the type of migratory event (birth, in-life, and death) is introduced. 

By applying this filtering procedure based on the time-window of interest, we reduce the number of Wikipedia biographies to be processed to $2,407$. 
We report in Table \ref{tab1} the distribution of work areas and disciplines present in this subset of Wikipedia biographies, which will be further processed as described in the next Section.

\subsubsection{Extracting migration footsteps}\label{subsec: extracting_trajectories}

For the purposes of our study, we are interested in identifying the different locations that were visited by the selected set of notable people during their lives and the year of their visits, in order to build a trajectory (made of multiple footsteps) for each notable person. This kind of information is not present in the Pantheon dataset and we therefore need to extract it automatically. The approach we adopt follows the one recently proposed with Ramble-On \cite{rambleon}, a text processing pipeline dealing with two main tasks: (i) the identification of predicates of migration and their arguments (i.e. the subject of the migration frame) in Wikipedia biography pages, and (ii) the recognition and classification of dates, places, and mentions. We focus our attention on migration processes because they are more likely to describe a motion action that resulted in a long time permanence in a new location. As a consequence, if the permanence in a specific location is long, it is more likely that the notable person had the time to provide his/her  cultural contribution there. While there have been recent attempts to automatically extract people's trajectories and an associated time period from Wikipedia biographical pages \cite{gergaud2016brief}, these rely on shallow NLP approaches based on the presence of keywords and geo-links in the pages. Furthermore, trajectories are coupled with time spans and not time points like in our study, leading to a lower granularity of the extracted information. In our case, the use of semantic parsing associated with a selection of predicates describing possible trajectories enables a very precise analysis of the resulting data.

To identify the predicates related to migration events, the Ramble-On application\footnote{Code available here: \url{https://github.com/dhfbk/rambleon}}  calls PIKES \cite{pikes}, a suite for NLP that extracts information from English plain texts, which in turn automatically assigns to each predicate a semantic frame based on the FrameNet classification \cite{framenet}. Also the arguments attached to each predicate are automatically labeled with semantic roles, relying on the frame-semantic parser Semafor \cite{semaphor} . 
To better distinguish migration predicates, we again follow the approach proposed in \cite{rambleon}, thus removing $16$ motion frames (e.g. \emph{Escaping}, \emph{Getting{\_}underway}, \emph{Touring}) out of $45$ because of the high number of false positives found during the identification. 
Hence, $29$ motion frames (e.g. \emph{Arriving}, \emph{Being{\_}employed}, \emph{Transfer}, \emph{Travel}) were used for the identification of  notable people's migration actions described in their biographies. 

Once a migration frame was identified in a sentence extracted from a Wikipedia biography page, three elements are required to be present in order to extract a migration trajectory: (i) the time/date of the motion, (ii) the traveler, and (iii) the destination. Again, the Ramble-On application selects only sentences satisfying these constraints, where the date, the notable person (or a reference to him/her) and the destination have been identified. With this approach, precision is favoured over recall, requiring that these three elements are explicitly mentioned in the same sentence. 

In Table \ref{tab2} we show as an example the snippet of a sentence identified as a movement. In this case the predicate ``moved'' is assigned to the frame ``Motion'' and it is selected as a migration frame. Then, the different arguments of the sentence are identified and labelled depending on their role in the sentence (e.g. the time, the traveller and the destination/place).

\begin{table*}[!t]
	\caption{Example of identification and classification of a sentence using Semaphor and FrameNet.}
	\label{tab2}
\centering
	\begin{tabular}{ccccc}
        \textbf{Snippet} & \textbf{Predicate} &  \textbf{Frame} & \textbf{Place} & \textbf{Time}\\
        \toprule
        "Paul moved to Chicago in 1934, &&&&\\ 
         where he continued to & moved & Motion & Chicago & in 1936  \\
         perform on radio."&&&&\\
        \bottomrule
    \end{tabular}
\end{table*}

Once the movements have been identified and the information about the date and the location have been extracted, Ramble-On geo-locates each word related to the identified destinations using  OpenStreetMap \textit{Nominatim}\footnote{\url{https://nominatim.openstreetmap.org/}}, a search engine for geo-referenced OpenStreetMap locations. Destinations that lack coordinates are discarded from the movements' list since they may be erroneously annotated as destinations. Besides, place and date of death are retrieved for each biography by Ramble-On using \textit{DBpedia}\footnote{\url{http://wiki.dbpedia.org/}}, where structured data about each notable person are stored.

In Table \ref{tab3} we present the information retrieved by processing the same biography as in Table \ref{tab2}. Together with the birth information retrieved from Pantheon, other important information such as date and place of death are extracted from \textit{DBpedia} as discussed above. The \textit{Migration 1} element, $M_1$, is the first additional trajectory retrieved by running the Ramble-On tool on the biography page. Merging the information retrieved about the inventor and musician Les Paul (see Table \ref{tab2}), we identify two jumps, i.e. two migratory events: (i) the first one from the birth location to \textit{Chicago}, and (ii) the second one from \textit{Chicago} to the death location.

\begin{table}[!ht]
	\caption{Results obtained by processing Les Paul's biography using the Ramble-On pipeline.}
	\label{tab3}
\centering
	\begin{tabular}{l|lll}
      & Birth & Movement 1 & Death\\
        \toprule
	    date & 19150000 & 19340000 & 20091231\\
	    place & Waukesha & Chicago & White Plains\\
	    latitude & 43.0117 & 41.8369 & 41.0400\\
	    longitude & -88.2317 & -87.6847 & -73.7786\\
	    predicate & null & moved & null\\
	    resource & dbpedia & FrameNet & dbpedia\\
	    place frame & null & @Goal & null\\
	    resource frame & Birth & Motion & Death\\
	    \bottomrule
    \end{tabular}
\end{table}

We refer to \cite{rambleon} for a detailed discussion on the performance of the extraction process. In brief, Ramble-On has a precision of $0.86$ in correctly identifying migration frames. As mentioned before, however, the strategy adopted to identify trajectories may penalize recall, failing to extract movements whose date or destination are mentioned implicitly or in two different sentences. In order to estimate the amount of locations visited (birth and death locations included) that we are not able to capture in our study, we manually annotate the trajectories in the biographies of $50$ notable people randomly sampled stratifying over the number of locations found. In this way, we have estimated that the recall of Ramble-on approach is equal to $0.59$.

\subsubsection{Dataset composition}
The processing of the $2,407$ biographies results in a set of $7,240$ locations connected with notable persons' trajectories. 
Among these, we consider the $4,028$ movements taking place in the 1900-1950 time-window. Each movement with the associated date and destination was then manually checked by comparing the extracted information with the source Wikipedia sentence, and corrected if necessary. Also the coordinates associated with the extracted locations by \textit{Nominatim} were manually checked, since the geographical information associated with trajectories is at the core of our migration model and possible errors must be minimized.
These are then collapsed to the nearest \textit{great city}, where we adopted as a definition of \textit{great cities} the list proposed in \cite{spatializing}. In their work, Reba \emph{et al.} \cite{spatializing} collected also precious historical demographic data for most of these cities, that we used to test our baseline for the migration model. More specifically,  geo-localized locations are merged based on a Voronoi tessellation of the Earth. In this framework each cell is built from a list of cities for which historical population data was available \cite{spatializing}. The space is built using the great-circle approximation to associate each identified location with the corresponding Voronoi cell. The distribution of the clustering process is reported in \add{Fig. SM7 of the Supplementary Materials (SM). In Fig. \ref{fig1} we present two examples of trajectories for Albert Einstein, the famous physicist, and Maria Montessori, the renowned  physician and educator. The arcs connects different locations where these two notable figures spent a part of their lives. The blue coloured side of an arc indicates the origin of the migration while the red one its destination. For example, the extraction well captures Einstein's movements from Zurich to Berlin and from Berlin to US. His first movement from Ulm, his home town, is missing since it happened before the beginning of 20th century. We also notice that his short period as visiting professor to Caltech is detected by Ramble-On. Similarly, Maria Montessori's experiences around Europe (i.e. Barcelona, Amsterdam, Vienna, Rome) are correctly identified. In contrast, we stress that, due to the lack of population data, her trips to Sri-Lanka are collapsed to cities in India. This shows how the collapsing process might impact the actual migration distribution.}

\begin{figure*}[!t]
\centering
    \includegraphics[width=0.9\linewidth]{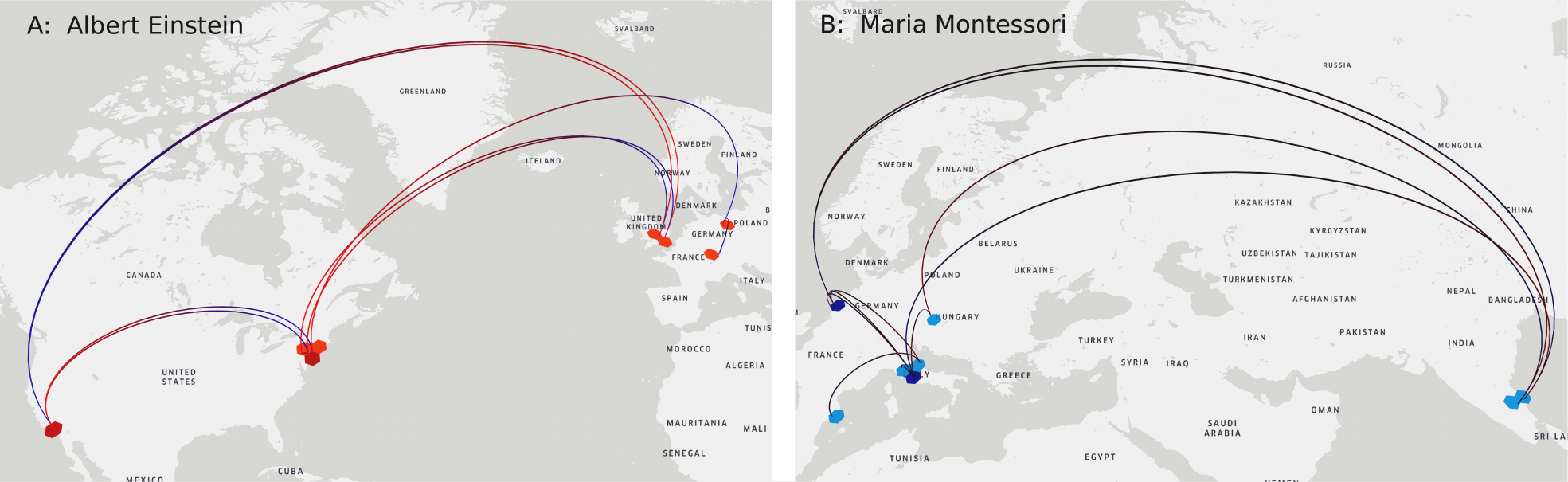}
\caption{Two examples of the trajectories obtained after the extraction. Panel A on the left shows the travels made by Albert Einstein while panel B those made by Maria Montessori.}
\label{fig1}
\end{figure*}

\begin{figure}
\centering
  \includegraphics[width=\linewidth]{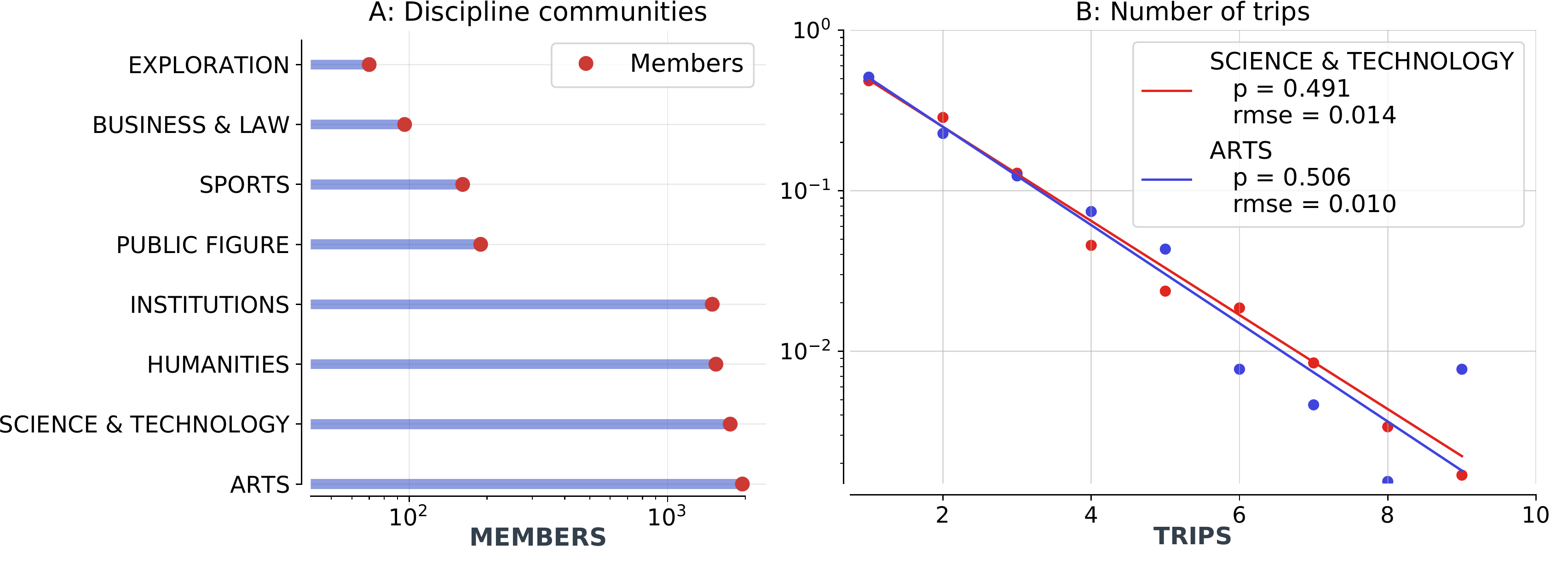}
\caption{\textbf{A}: Size of disciplinary communities present in the dataset. From the bar-plot it is evident the difference in the number of notable individuals between arts, science\&technology, humanities, institutions and public figures, sports, business\&law, exploration. \textbf{B}: Distribution of the number of trajectories per person in the set of notable people. The lines represent the geometric fit to the distributions of the two most numerous disciplines, i.e. \textit{Arts} and \textit{Science and Technology}. Both  distributions can be described in terms of a geometric distribution with parameter $p\sim \frac{1}{2}$.}
\label{fig2}
\end{figure}

The merging step results in a set of $629$ different cities visited by our notable people during their lifetime.
Figure \ref{fig2}.A shows the members of our set of notable individuals listed by discipline (as labelled by \cite{pantheon_yu}); while Figure \ref{fig2}.B shows the distribution of different visited cities for the top two discipline communities, namely "Arts" and "Science and Technology". The colored dots represent the data while the lines the geometrical fit to the data. Both the distributions can be described using a geometrical distribution with parameter $p$, representing the probability of successfully settling in a city, $p\sim\frac{1}{2}$.

\subsection{Cultural network and migration modeling}\label{subsec: network_construction}

In our framework, we assume that a \add{culturally} notable person living in a place for a certain period of time contributed in some way to make such place a cultural attractor for other people interested in cultural innovation and development. At the same time, when a culturally notable person was moving from one place to another, s/he linked the cultural relations s/he had and the work s/he did in the first place with the relations and work in the second place. In this way, each movement creates a cultural connection between two different places around the world. Depending on the number of notable people moving from one place to another, we can add a weight to the links of this cultural network. Thus, the nodes correspond to the different cities visited by our set of notable people, while the edges are cultural links, built by interpersonal relations together with the cultural contamination a person brings with herself/himself while migrating from one place to another, weighted by the number of occurrences. 

\begin{figure*}[!t]
\centering
\includegraphics[width=\linewidth]{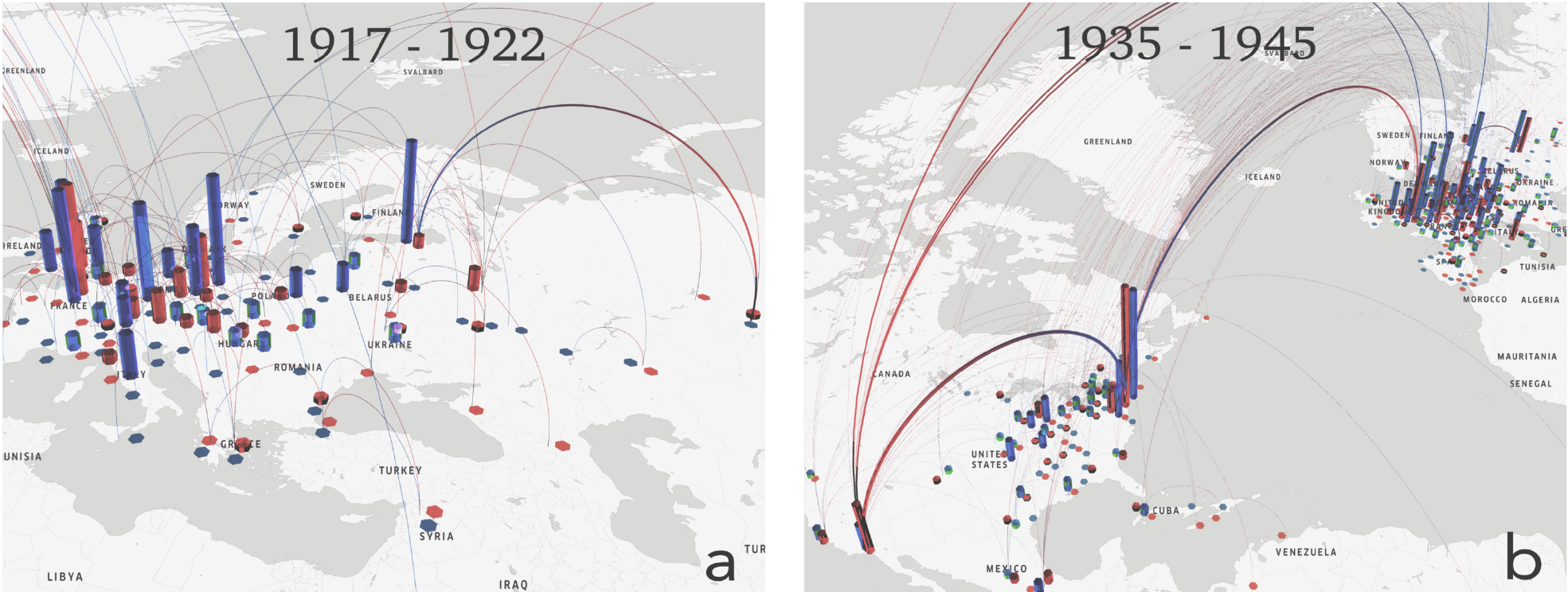}
\caption{A representation of the spatial directed network encoding the mobility information of the chosen set of notable people during the first half of the 20th century. We use the blue hexagon to indicate the source point and the red to indicate the destination of the migratory jump. In this figure, we present the network structure considering the migratory events during two different time snapshots. Figure \ref{fig3}.a on the left shows the escaping and exiling from Saint Petersburg of the aristocratic families during the years of the red revolution. Similarly, Figure \ref{fig3}.b on the left shows the migration flows from Europe to North America during World War 2, due to the fascist regimes and persecutions.}
\label{fig3}
\end{figure*}

In Figure \ref{fig3} we show a representation of this weighted directed network, where the weight is the percentage of notable people migrating from a location to their destination. In the left panel and right panel we show a sector of the network considering different time frames. On the left, we restrict our focus on a 3-year time-window centering the map to highlight the connections between Eastern Europe and Asia (Union of Soviet Socialist Republics in particular) during the October revolution in 1917. It is interesting to notice that the structure of the network nicely catches the known phenomenon of Siberian exile of aristocratic families and important political personalities such as representatives of the previous tsarists' power and relevant persons not aligned with the current regime. In this example, we are capturing the movements of people that were forced to migrate to specific locations. Hence, our notion of cultural attractor both includes locations to which people moved voluntarily and locations to which culturally notable people were forced to move. In particular, the number of forced movements (i.e. people sent to concentration camps and imprisoned people) is 36 out of 3474 identified movements. Top right panel shows the strong migratory flux of intellectuals from Europe to the US during the Second World War. 

As previously said, we are interested in modeling the mobility of culturally relevant figures and in investigating the factors playing a role in their migration patterns. To this end, we modify the radiation model to take into account the role played by the city size (i.e. proxy for job opportunities) as well as the ones played by the different disciplines and by the number of notable people (i.e. proxies for cultural opportunities). The radiation model \cite{radiation} describes the mobility of people seeking job opportunities in terms of job openings per number of inhabitants. The model is developed in the framework of network theory since it treats cities as nodes of a completely connected weighted network. Specifically, the radiation model describes the human mobility behavior at long distances, e.g. at the country or global scale, better than other often used models (e.g. gravity model) \cite{masucci_graVSrad}. It is also important to notice that it undershoots the real flows. Moreover, its performances are dependent on the structure of the system even though it directly accounts for variations of the population between the source and the destination of a migration, i.e. the less population you have between two cities the more probable is to migrate from one to the other.

Simini \emph{et al.} \cite{radiation} show that, using this formulation of the problem, the flow of people between cities only depends on the population of the two cities (namely $m_{i}$ and $n_{i}$), and the population living in the circle of radius $r_{ij}$ is equal to the distance between the two cities (namely $s_{ij}$). The relation can be summarized in a simple and parameter-free equation. We report here the formula for the probability to move from city $i$ to city $j$:
\begin{equation}
P_{ij}  = P_i\frac{n_j}{(m_i+s_{ij})(m_i+n_j+s_{ij})},
\label{eq:radiation}
\end{equation}
where $P_i$ is the normalization coefficient for city $i$ that ensures that $P_{ij}$ is the probability of moving from $i$ to every city ($i$ included): $P_{i}  = \sum_{j\in N}\frac{(m_i+s_{ij})(m_i+n_j+s_{ij})}{n_j}$, where $N$ is the set of all nodes present in the network.
In our work we make use of these concepts to model in a similar way the mobility of culturally relevant people. \add{In particular, inspired by multidimensional network theory and it recent applications in modeling human mobility  \cite{DeDomenico_MN,ratti,ferretti}, we propose a multilevel approach to cultural mobility. In this framework, every} cultural discipline works as a separate system described by a cultural radiation model. Formally, a level is a fully connected weighted and directed network in which the nodes are the cities visited by all the notables of a specific discipline and the links represent the probability of migrating from a city to a different one. Each node has also a link pointing to itself, representing the probability of remaining in the same city instead of moving to a different place. The different levels, $l\in L$ where $L$ is the set of all disciplines, do not interact with each other but their contribution to the overall migratory exploration sums up. Each level contributes to the global migration model with a factor proportional to the share of notable people the discipline has, $NS$. Thus, the probability of this multilevel migratory network can be described by
\begin{equation}
P_{ij}  = \sum_{\forall l\in L}NS_lP_{i_l}\frac{n_{j_l}}{(m_{i_l}+s_{ij_l})(m_{i_l}+n_{j_l}+s_{ij_l})},
\end{equation}
where $NS_l$ is the notable share of discipline $l\in L$, $m_{i_l}$ and $n_{j_l}$ are the population of locations $i$ and $j$ respectively in the discipline level $l$, and $s_{ij_l}$ is equivalent to $s_{ij}$ for the specific level $l$. In a similar way, the generalized $P_{i}$ normalizes $P_{ij}$ to a probability following the idea of equation (\ref{eq:radiation}): 
\begin{equation}
P_{i_l}  = \sum_{j\in N}\frac{(m_{i_l}+s_{ij_l})(m_{i_l}+n_{j_l}+s_{ij_l})}{n_{j_l} }
\end{equation}

Starting from these two equations, we propose different implementations of the radiation model. In particular, we stress that the radiation model introduces the concept of attractiveness of a city based on the number of job opportunities that a city can provide. The assumption that this number is directly proportional to the number of people living in a city \cite{radiation} directly connects the concept of attractiveness with the population size of a city. Here, with similar assumptions we propose different formulations of the model based on different possible ways of modeling cultural attractiveness. In particular, we use, as possible alternatives to the standard formulation of the model, the number of notable people that visited a city and its combination with the population of the city.
In the case of notable people that visited a city, we count this quantity considering all the visits during the whole time-window. As a consequence, we model cultural dynamics of individuals within this period of observation considering the effects of the notable people distribution as a constant feature of our model, as we do for general population, which is not updated after every step of the dynamics. This is equivalent to assume that a single step of the dynamics has a latency larger than the size of the considered time-window in affecting the importance of cultural attractors. Moreover, counting notable people in this way also relies on the simplifying assumption that they all equally contributed to the importance of a city from a cultural perspective. More realistic modeling will require a \textit{relevance score} based on the historical relevance of the notable people.

We test each of these three possible definitions, namely the standard one based on population size, the one based only on the number of notable people, and the combination of the two, using both a single level formulation of the radiation model and a multilevel formulation.
Using this probabilistic model, our aim is to understand if the radiation model abstraction can be used to describe the level of exploration of the historically and culturally relevant figures (namely, the radius of gyration of each notable people, the number of different cities visited, and the distance distribution of the migration jumps) and which formulation better captures these properties.
In the next Section, we discuss first the general information that can be obtained by analyzing the system in terms of network theory metrics and then the comparison between the different formulations proposed.
\vspace{-0.3em}

\section{Results} \label{results}
\subsection{Properties of the migration network}
\begin{figure}
\centering
  \includegraphics[width=\linewidth]{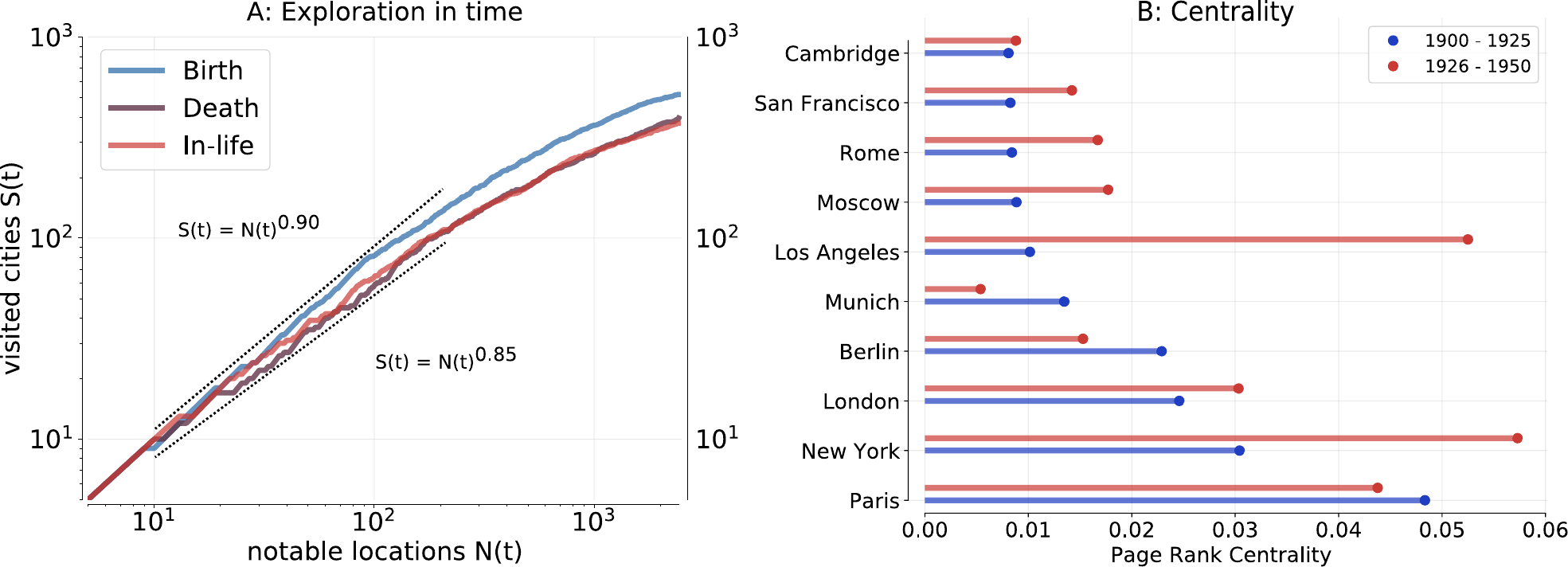}
\caption{\textbf{A}: Number of different visited cities as a function of the notable birth/death/in-life visited locations. In the case of in-life migrations the number of visited locations does not necessarily corresponds to the number of notable figures in the dataset, but it represents the different cities visited in time as a function of the number of trips made by the notable figures. \textbf{B}: Page-rank centrality for the global mobility network evaluated in two different time-windows. In blue we report the centrality values for the 1900--1925 window, while in red the centrality in the window 1926--1950.}
\label{fig4}
\end{figure}

One of the most interesting characteristics of cultural migration patterns is the tendency of notable figures to explore different cities. To study this property, we can define $S(t)$ as the number of different cities and $N(t)$ as the number of notable people's birth locations, the number of their death locations, and the number of their jumps during the selected time-window, as displayed in Figure \ref{fig4}.A. by the curves for \textit{Birth}, \textit{Death} and \textit{In-life}  respectively. The growth of $S(t)$ is modeled as a function of $N(t)$ using a Heap's law $S(t)=N(t)^{\alpha}$. Our result is consistent with the estimate of the parameter $\alpha$ for the \textit{Birth} curve obtained by Schich \emph{et al.} in \cite{culture}. A similar result is obtained also for the \textit{Death} and \textit{In-life} curves representing the growth of the location for which the exponent, $\alpha = 0.85$, suggests a tendency to migrate to a smaller number of cities with respect to the number of different cities where notable individuals were born. This finding may be interpreted as a general and global tendency of notable figures to migrate to a more {culturally renowned} subset of cities with respect to all the possible available locations.

Focusing on the migration jumps that notable people made during their life, we can study the most central cities both from a global and discipline-based perspective.
Here, we use the Page-Rank centrality \cite{brin1998} to measure the importance in terms of the number of incoming links that point to a city and the relative importance of the cities from which these links are coming. In Figure \ref{fig4}.B, we measure Page-Rank centrality for two different time-windows, namely 1900-1925 and 1926-1950, to show the structural changes of the network during the first half of the 20th century. It is interesting to notice how the development of the film industry in Los Angeles attracted several figures to the city. It is also worth noticing how, due to the Second World War (WW2), Berlin loses positions in the ranking of the \textit{more central} cities. In Figure SM4, we highlight the specific effect of WW2, evaluating Page-Rank centrality before and after the rise of the Nazism regime in Germany, showing the overall loss in cultural centrality for most of the European cities.

\begin{figure}
\centering
  \includegraphics[width=\linewidth]{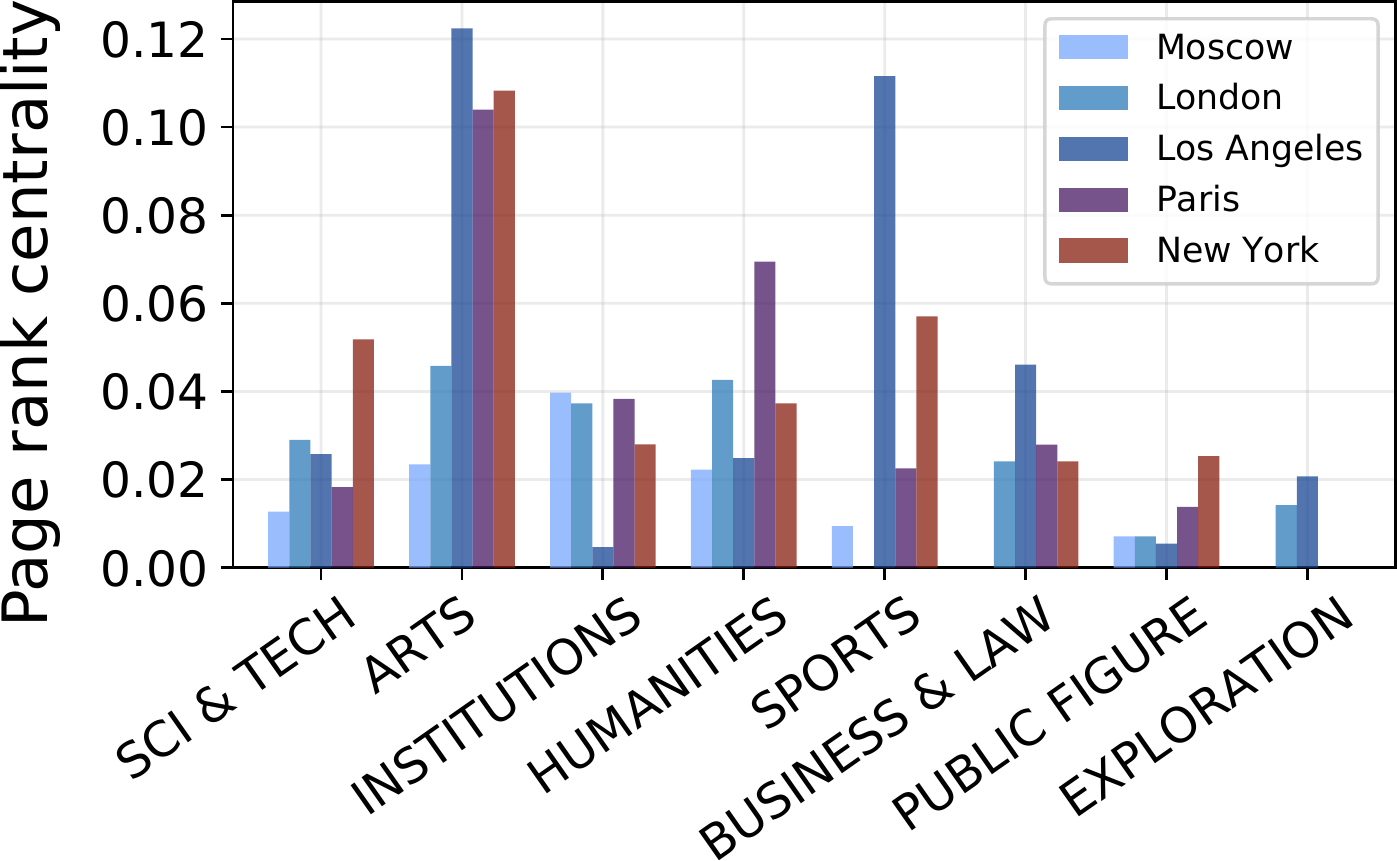}
\caption{Page rank centrality per discipline for the five most visited cities. This figures shows the different importance of cities in terms of discipline specific attractiveness.}
\label{fig5}
\end{figure}

We also evaluated Page-Rank centrality for the sub-network built only by considering the migration jumps of the four top disciplines in those years. Figure \ref{fig5} shows how notable people from different disciplines migrated to different cities, suggesting that the cultural centrality of a city depends on its cultural characteristics, e.g. Los Angeles and the film industry. \add{This, indeed, results in Los Angeles being a central node for Arts (in particular for \textit{actors}) and Sports, but a more peripheral one for Institutions, having their more central nodes in capital cities such as London, Paris and Moscow. In Fig. SM3 we performed a discipline Page-Rank analysis for two time windows as in Fig. \ref{fig4}, showing that an important change is present also at the discipline level. An example is given by the dramatic change (i.e. a decrease) in the centrality of Berlin for the scientific community before and after 1933.}

\subsection{Cultural attractiveness in a multilevel radiation model}
With the results obtained so far we aim at modeling notable figures' migratory patterns to better understand what is the process driving the choice of the location where to migrate.
The radiation model proposed in \cite{radiation} finds the motivating factor of mobility for those seeking a job in the number of opportunities a city can provide. Following this idea and using the results obtained by Simini \emph{et al.} \cite{radiation}, we propose a similar approach to understand if such a model can catch the main factors of cultural mobility. 
We assume that the number of opportunities that are available in the selected time-window is directly proportional to the number of notable figures that lived in a city during the same time interval.
In particular, we explore the following different configurations: 
\begin{itemize}
    \item cultural opportunities are uniformly distributed among cities;
    \item cultural opportunities are proportional to the population of a city;
    \item cultural opportunities of a city are directly proportional to the number of notable figures that lived in that city;
    \item cultural opportunities are directly proportional both to the population of a city and to the number of notable figures that lived in that city.
\end{itemize}
In addition, we also want to check whether cultural opportunities depend on the discipline a notable individual is part of. We study all these possibilities using the formulation proposed in Section \ref{subsec: network_construction}. 

To find which model better describes the historical mobility of the first half of 20th century, we simulate  mobility using a set of walkers that can move following a radiation model over the cultural network, based on different equations depending on the model we are simulating. Thus, depending on the selected configuration, we are using respectively (i) a random walker dynamics' model over the network, (ii) a standard (population-related) radiation model, and (iii) an implementation of the radiation model that specifically considers the cultural opportunities as a subset of job opportunities not necessarily driven by the same factors. 
The starting location of each simulated notable individual is chosen based on a random choice weighted by the population size of each city.

To compare the models we analyze their impact on predicting (i) the number of different cities visited, as a proxy of the availability to explore different and new destinations, (ii) the radius of gyration of each notable figure simulated, and (iii) the distribution of the length of the migration jumps. We report in Table \ref{tab4} the results obtained for five representative models: (i) the random walker model, (ii) the notable-based jump probability on a single-level structure, (iii) the notable-based jump probability on a multilevel structure, (iv) the population-based jump probability on a multilevel structure, and (v) the mixed population-notable-based jump probability on a multilevel structure. Results for the other models are reported in the SM, specifically in Tables SM1, Table SM2, and Table SM3.

    \begin{table*}[!t]
    \caption{Models' performances. Results obtained for the five models on predicting the number of destinations, the radius of gyration, and the distribution of the length of the migration jumps. The metrics used are the adjusted-$R^2$, the Pearson correlation coefficient, $\rho$, between models and data, the Kullback-Leibler distance (K-L dist), and the first Wasserstein distance (Wasserstein dist).}
    \centering
    \resizebox{\textwidth}{!}{
    	\begin{tabular}{rcccc}
            \textbf{Model} &  $\mathbf{adj-R^2}$ &   \textbf{Pearson }$\mathbf{\rho}$ & \textbf{K-L dist} & \textbf{Wass. dist}\\
            \toprule
            \textbf{\textit{Radius of gyration}}&&&&\\
            \textbf{pop-notable-multilevel}&$0.2414\pm0.0027$&$0.962^{***}$&$\mathbf{0.00554\pm0.00004}$&$\mathbf{0.000100\pm 2e-7}$\\
            pop-multilevel&$-0.2004\pm0.0034$&$0.953^{***}$&$0.00655\pm0.00005$&$0.000125\pm1.e-7$\\
            notable-multilevel&$-0.6849\pm0.0041$&$0.947^{***}$&$0.00836\pm0.00005$&$0.000139\pm1e-7$\\
            notable-singlelevel&$-1.0249\pm0.0048$&$0.923^{***}$&$0.01006\pm0.00006$&$0.000143\pm1e-7$\\
            random-singlelevel&$-2.2673\pm0.0054$&$0.886^{***}$&$0.01559\pm0.00009$&$0.000173\pm1e-7$\\
            \midrule
            \textbf{\textit{Different destinations}}&&&\\
            \textbf{pop-notable-multilevel}&$0.9547\pm0.0004$&$0.978^{***}$&$0.0649\pm	0.002$&$\mathbf{0.0150\pm0.0001}$\\
            pop-multilevel&$0.9612\pm0.0004$&$0.981^{***}$&$\mathbf{0.0561\pm0.001}$&$0.0154\pm0.0001$\\
            notable-multilevel&$0.9619\pm0.0003$&$0.982^{***}$&$0.0570\pm0.001$&$0.0155\pm0.0001$\\
            notable-singlelevel&$0.9624\pm0.0003$&$0.982^{***}$&$0.0623\pm0.002$&$0.0159\pm0.0001$\\
            random-singlelevel&$0.9606\pm0.0004$&$0.982^{***}$&$0.0724\pm0.002$&$0.0163\pm0.0001$\\
            \midrule
            \textbf{\textit{Length of migration jumps}}&\\
            \textbf{pop-notable-multilevel}&$0.5104\pm0.0019$&$0.982^{***}$&$\mathbf{0.00533\pm0.00005}$&$\mathbf{0.000080\pm1e-7}$\\
            pop-multilevel&$0.2249\pm0.0023$&$0.974^{***}$&$0.00686\pm0.00005$&$0.000099\pm1e-7$\\
            notable-multilevel&$-0.0640\pm0.0029$&$0.967^{***}$&$0.00795\pm0.00005$&$0.000109\pm1e-7$\\
            notable-singlelevel&$-0.2192\pm0.0029$&$0.962^{***}$&$0.00790\pm0.00006$&$0.000112\pm1e-7$\\
            random-singlelevel&$-0.8313\pm0.0034$&$0.947^{***}$&$0.01265\pm0.00006$&$0.000131\pm1e-7$\\
            \bottomrule
        \end{tabular}}
    	\label{tab4}
    \end{table*}

Among the different possibilities tested, we found that a modeling approach considering cultural attractors as a product of both job-opportunities and cultural interests, by means of the population number and the effective number of notable people migrated in the time-window under investigation, better captures key features of notables' mobility. Moreover, we stress that the model treating different disciplines as different dynamics outperforms the single-level models in terms of Kullback-Leibler divergence \cite{kullback1951} and first Wasserstein distance \cite{olkin1982}.

\begin{figure}
    \centering
    \includegraphics[width=\linewidth]{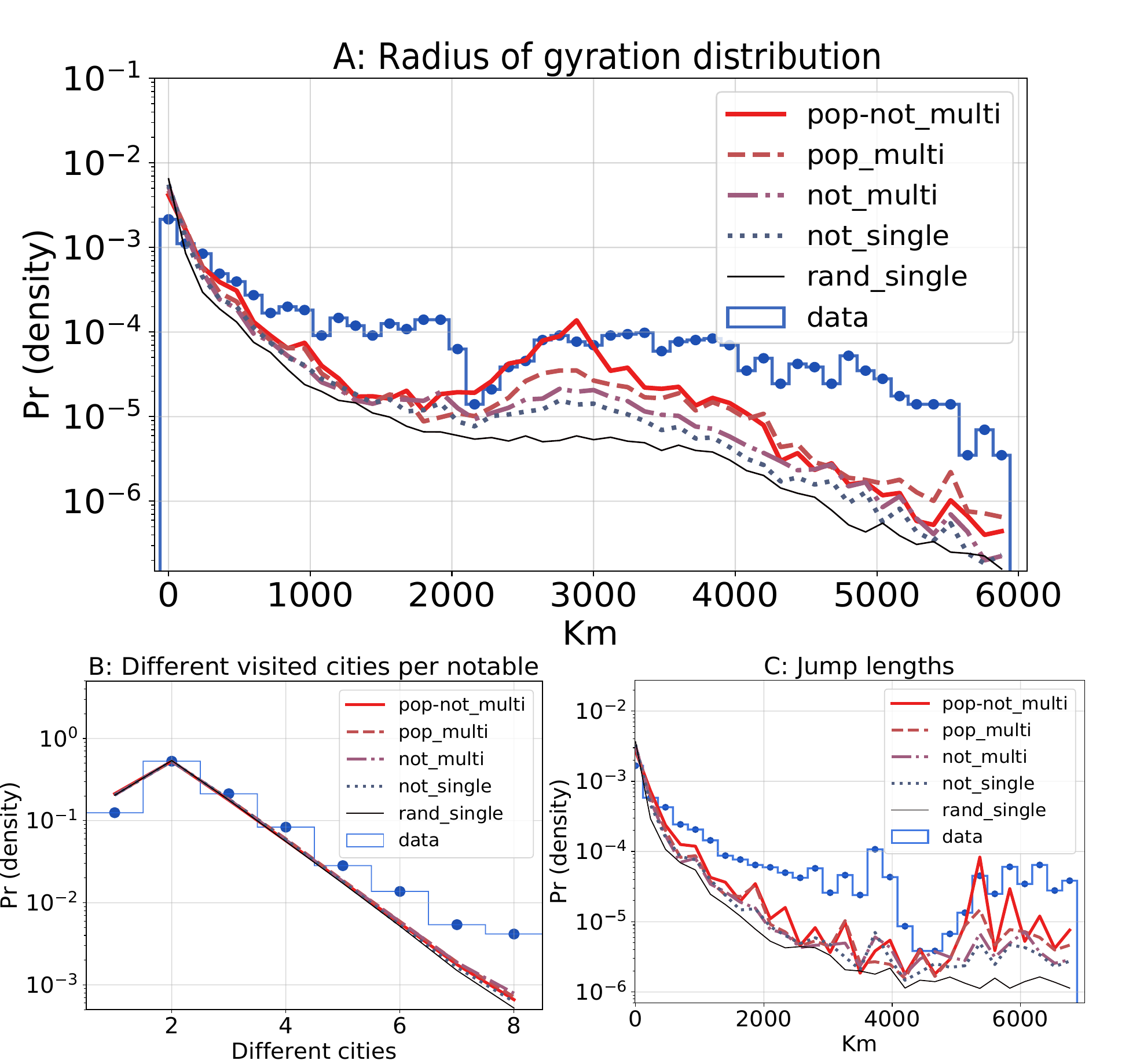}
    \caption{\textbf{A}: The distribution of the radius of gyration per notable person; \textbf{B}: The distributions of the data and the five different models for the number of different cities visited per notable person; \textbf{C}: The distribution of the lengths of the all the jumps simulated for the different model (lines) and for the data (stepped density histogram).}
    \label{fig6}
\end{figure}

These quantities are measured after simulating the mobility for $2,000$ notable people, whose number of migration jumps was randomly sampled from a geometric distribution with parameter $p\sim0.5$. We repeated the simulation for 500 times to estimate the stability and the standard error of the different metrics.
Figure \ref{fig6}.A shows the distribution of the mean radius of gyration of the $500$ simulations against the data (blue-stepped line). Figure \ref{fig6}.B shows the distribution of different cities visited by a notable person during her/his lifetime, while the distribution presented in Figure \ref{fig6}.C shows the probability of jumping to a destination that is at a specific distance from the origin. These distributions are highly dependent on the geographical distance. Short distance migrations are largely privileged while at $\sim 2500 km$ and $\sim 5500 km$ respectively a second smaller peak appears capturing the overseas migrations (across the Atlantic Ocean) mainly from Europe to US. We stress that the effect of slightly underestimating the log-distance trips, which also affects the radius-of-gyration distribution, has been proven to be a structural feature of the radiation model under irregular geographical configurations such as those imposed by oceans \cite{masucci_graVSrad}.

\section{Discussion and Conclusions} \label{discussion}
A complex question has been posed in \cite{culture}, i.e. whether it is possible to describe the dynamical properties of the cultural migration phenomenon. Starting from their idea of using network theory to tackle the problem, we make some further steps in understanding this issue. First of all, we use NLP tools to capture a more detailed representation of the lives of \add{historically} notable people that can be considered as cultural developers or important actors in the evolutionary process of culture. Our approach gathers information not only from the birth and death events but also from in-life migratory events, enabling us to study in a more detailed way the cultural migration processes and to include in our model the years in which a person is professionally more active. 
Indeed, there is a difference between the birth location of a person and the locations s/he migrates to or where s/he decide to spend her/his last years. The first birth location is not determined by a decision of the born individual, while the in-life migrations and the death places are more likely to be chosen following some precise interests and motivations. Using our data we are able to capture this difference and quantify the contribution of the exploration level of notable people during these phases of life.

Moreover, we focus our attention on understanding the main features that drive this kind of mobility. \add{Our results provide evidence that the mobility of historically and culturally notable individuals is best described by simultaneously} considering three different factors: (i) the population of a city, as a proxy of economic wealth and generic job opportunities; (ii) the number of culturally notable people that spent some time of their lives there, as a proxy of the attractive role played by this city as a cultural hub and of the proneness of this city to invest in culture; and (iii) the discipline a culturally notable person is working in, as a proxy both of the interest a city has in investing on a specific cultural area and the tendency of people, interested or working on a given discipline, to follow notable ones from the same discipline. \add{The solution proposed in our work represent a functional integration, in a quantitative theoretical model, of these components.}

It is also worth highlighting some limitations of our work. First of all, \add{we rely on Wikipedia as data source, which shows a clear bias towards the Western culture and male figures. This limitation is even more relevant since we focus only on pages written in English}. Then, we consider a specific time-window in the cultural history, i.e. the first half of the 20th century. So, our results may be dependent on the time chosen (for example, we may observe  different behaviors during wider time-windows) and on the small available dataset of \add{historically and} culturally notable personalities. Besides, while our data on trajectories were extracted automatically but manually revised, we estimate a recall of our information extraction pipeline of $0.59$, as pointed out in Section \ref{subsec: extracting_trajectories}. This implies that some migration destinations that are mentioned in the Wikipedia biographies should be added (e.g. by improving the extraction performances of Ramble-On) in order to make our set of data even richer. A richer dataset will also help in stabilizing, constructing and precisely characterizing the structure of the network. However, we also stress that the extension to the radiation model proposed here only uses the visited locations and not the migration timelines of notables. Thus, the present recall level in extracting complete timelines does not directly affect the dynamical structure of our model. Thanks to these considerations, it is also interesting to discuss the specific limitations of the mobility model proposed. In particular, while the fit of the number of trips might depend on the recall limitations discussed above and thus affect the number of different visited locations, this does not explain the systematical underestimation of the probability of \textit{high-distance jumps} and \textit{high-radius of gyration}. We expect the geographical constraints (as discussed in Section \ref{subsec: network_construction}) and the specific time-window we selected (e.g. WW2 forced to migrate many notable people whose choice was biased towards US) to be two determinants of these discrepancies.

Overall, our results open interesting possibilities on further investigating the historical role played by places and cities in attracting culturally relevant figures as well as on better analyzing the level of contribution of each of the factors identified by our approach (i.e. city's population, number of intellectuals living in the city, and strength of a specific cultural discipline in the city).\add{Similarly, changing the perspective, it will become possible to quantify the impact of cultural communities on local well-being, helping our understanding on how individuals from similar or different disciplines combine and collaborate to seed the vital growth of cities' economies.}

% \section{Conclusion} \label{conclusions}

%\begin{appendices}
%\section*{Appendix}
%\end{appendices}
    
    %%%%%%%%%%%%%%%%%%%%%%%%%%%%%%%%%%%%%%%%%%%%%%
    %%                                          %%
    %% Backmatter begins here                   %%
    %%                                          %%
    %%%%%%%%%%%%%%%%%%%%%%%%%%%%%%%%%%%%%%%%%%%%%%
    
    \begin{backmatter}
    \section*{Availability of data and material}
    The data used in this work are available at:
    \add{\url{https://doi.org/10.7910/DVN/PJS21L} or\newline}
    \url{https://figshare.com/articles/Following_the_footsteps_of_giants/7352987}. \newline
    The code used to extract the destinations from the Wikipedia biographies is publicly released at \url{https://github.com/dhfbk/rambleon}.

    \section*{Competing interests}
     The authors declare no competing financial or non-financial interests.
    
    \section*{Author's contributions}
     All authors conceptualized the project. L.L acquired and cleaned the data, performed the investigation, the statistical analyses and drafted the original manuscript. All authors contributed revising the manuscript and gave final approval for publication.
     
    \section*{Funding}
    No funding supported our research.
    
    \section*{Abbreviations}
    NLP: Natural Language Processing; WW2: World War 2; US: United States; SM: Supplementary Material.

    %%  After submission of the .TEX file,                     %%
    %%  you will be prompted to submit your .BBL file.         %%
    
    % if your bibliography is in bibtex format, use those commands:
    \bibliographystyle{bmc-mathphys} % Style BST file (bmc-mathphys, vancouver, spbasic).
    \bibliography{bmc_article}      % Bibliography file (usually '*.bib' )

\end{backmatter}
\end{document}